\documentclass[final,12pt,figuresleft, pdftex]{clear2024} 

\usepackage[utf8]{inputenc} 
\usepackage[T1]{fontenc}    
\usepackage{hyperref}       
\usepackage{url}            
\usepackage{booktabs}       
\usepackage{amsfonts}       
\usepackage{nicefrac}       
\usepackage{microtype}      
\usepackage{xcolor}         
\usepackage{listings}    
\usepackage{amsmath}
\usepackage{adjustbox}
\usepackage{array}
\usepackage{rotating}
\usepackage{graphicx}

\newcolumntype{P}[1]{>{\centering\arraybackslash}p{#1}}
\newcolumntype{M}[1]{>{\centering\arraybackslash}m{#1}}

\definecolor{dkgreen}{rgb}{0,0.6,0}
\definecolor{gray}{rgb}{0.5,0.5,0.5}
\definecolor{mauve}{rgb}{0.58,0,0.82}

\newtheorem{example2}{Example}

\usepackage[skip=5pt]{caption}

\usepackage{csquotes}

\title[The PetShop Dataset]{The PetShop Dataset — Finding Causes of Performance Issues across Microservices}
\usepackage{times}

\clearauthor{%
 \Name{Mila Hardt} \Email{milaha@amazon.com}\\
 \addr Amazon
 \AND
 \Name{William R. Orchard}\thanks{Work done while interning at Amazon Research in Tübingen, Germany} \Email{william.orchard@cruk.cam.ac.uk}\\
 \addr University of Cambridge
 \AND
 \Name{Patrick Blöbaum} \Email{bloebp@amazon.com}\\
 \addr Amazon
 \AND
 \Name{Elke Kirschbaum} \Email{elkeki@amazon.com}\\
 \addr Amazon
 \AND
 \Name{Shiva Prasad Kasiviswanathan} \Email{kasivisw@gmail.com}\\
 \addr Amazon 
}

\usepackage{xr}
\makeatletter

\newcommand*{\addFileDependency}[1]{
\typeout{(#1)}
\@addtofilelist{#1}
\IfFileExists{#1}{}{\typeout{No file #1.}}
}\makeatother

\newcommand*{\myexternaldocument}[1]{%
\externaldocument{#1}%
\addFileDependency{#1.tex}%
\addFileDependency{#1.aux}%
}

\myexternaldocument{supplementary_material}

\begin{document}
\maketitle

\begin{abstract}
Identifying root causes for unexpected or undesirable behavior in complex systems is a prevalent challenge. This issue becomes especially crucial in modern cloud applications that employ numerous microservices. Although the machine learning and systems research communities have proposed various techniques to tackle this problem,
there is currently a lack of standardized datasets for quantitative benchmarking. Consequently, research groups are compelled to create their own datasets for experimentation.
This paper introduces a dataset specifically designed for evaluating root cause analyses in microservice-based applications. The dataset encompasses latency, requests, and availability metrics emitted in 5-minute intervals from a distributed application.  In addition to normal operation metrics, the dataset includes 68 injected performance issues, which increase latency and reduce availability throughout the system. We showcase how this dataset can be used to evaluate the accuracy of a variety of methods spanning different causal and non-causal characterisations of the root cause analysis problem.  We hope the new dataset, available at \url{https://github.com/amazon-science/petshop-root-cause-analysis}, enables further development of techniques in this important area. 
\end{abstract}

\section{Introduction} \label{sec:intro}
 Identifying the origin of unexpected or undesired behavior of a system, also called root cause analysis (RCA), is of extreme relevance in various disciplines.  For example, in manufacturing pipelines, if certain parts exceed failure rate thresholds during quality tests, it is crucial to identify and resolve the root cause to avoid financial losses and reputational damage. The timely resolution of such issues is vital as prolonged downtime can result in substantial monetary losses and erode customer trust. Likewise, for online retailers, any decrease in website availability or prolonged page load times directly impacts revenue and customer confidence, underscoring the urgency of RCA in restoring normal operations.
 Identifying the root cause of such issues, however, can be extremely cumbersome and time-consuming, particularly in complex applications composed of tens or hundreds of microservices. Such microservice architectures have become extremely popular in recent years, since decomposing an application into different microservices that communicate through agreed upon application programming interfaces (APIs) enables clear ownership and rapid development. Additionally, the most suitable hardware can be chosen  for each component and scaled up or down independently. These are great advantages compared to a monolithic application architecture. However, when problems arise, identifying the cause in these complex systems is challenging and requires not just knowledge of the individual components, but also about the interaction between the components. Oncall engineers may need to look over hundreds of metrics, dig in terabytes of logs, ping people from other teams responsible for various components, before they obtain a clear picture of what went wrong. 

To reduce the mean time to resolution, numerous methods have been proposed for automating RCA using available system data. Some of these methods are specifically tailored for RCA in microservice-based applications  \citep[e.g.][]{rca_causeinfer,rca_microscope,rca_automap,rca_sage,rca_MicroHECL,rca_ikram_neurips,rca_circa}, while others are designed for a broader class of use cases \citep[e.g.][]{Budhathoki2022}.
Among these~\citet{rca_ikram_neurips} has released a dataset based on the Sock-shop application to evaluate such methods. 
In this paper we introduce a similar dataset also based on microservice architecture with issues injected. However, the datasets differ in a few crucial aspects: our dataset is collected from a microservice based application which contains more than 3x as many components, we inject more issues, cover more types of issues beyond the two considered in the Sock-shop dataset, and include multiple different traffic generation patterns. This allows us to consider a greater diversity of scenarios to study which techniques work well in, which we show in our experimental section.

 The dataset encompasses latency, requests, and availability metrics, gathered from a distributed application comprising 41 components, including databases, load balancers, queues, storage systems, and containerized microservices. In addition to normal operation metrics, the dataset includes 68 injected performance issues, such as request overload, memory leaks, CPU hog, and misconfigurations, which increase latency and reduce availability throughout the system. The metrics are annotated with the corresponding issues, serving as ground truth for the analysis. Interestingly, methods that performed well on the Sock-shop data struggle on our dataset. This illustrates the value of having this additional dataset to broaden our understanding of RCA methods, and motivates future research in robust data-efficient RCA methods. This dataset addresses the concern of biased reporting by avoiding selective focus on issues where the developed method demonstrates strong performance. 

Our goal is therefore to accelerate research on RCA methods by allowing researchers to focus on the development of new methods instead of generating data for their evaluation. The dataset is publicly available, alongside code for running the experiments described in Section~\ref{sec:experiments} below, at \url{https://github.com/amazon-science/petshop-root-cause-analysis}. 
The dataset is designed with an easily extendable data format and accompanying tooling, encouraging broader participation and contribution from others.

 The remainder of the paper is organized as follows: next, we present an overview of the causal formalization of root-cause-analysis, laying out the different approaches to answering the question, `what is a root cause?' In Sec.~\ref{sec:related_work} we describe some of the methods which have been developed for RCA in the context of microservice-based applications and the datasets available to evaluating them. We describe our dataset in Sec.~\ref{sec:description} and compare  existing methods on it in Sec.~\ref{sec:experiments}. 
 After disclosing limitations in Sec.~\ref{sec:limitations} we conclude in Sec.~\ref{sec:conclusion}.
\section{Background} \label{sec:background}
In this section we give a brief survey of different approaches to defining a `root cause'. Broadly, methods can be categorized according to what layer of the Causal Hierarchy~\citep{PearlMackenzie18,Bareinboim2022-BAROPH-2} they pose the RCA problem. The causal hierachy delineates a strict hierarchy wherein knowledge at each layer enables reasoning about different classes of causal concepts. Layer 1, $\mathcal{L}_{1}$, is associational, layer 2, $\mathcal{L}_{2}$, is interventional, and $\mathcal{L}_{3}$ is counterfactual. The problem setup is common: the value $x_{n}$ of target variable $X_{n}$ has been flagged as anomalous. With jointly observed values $(x_{1},...,x_{n})$ of variables $(X_{1},...,X_{n})$, we must find the root cause(s), among these variables, of the anomaly $x_{n}$.

At $\mathcal{L}_{1}$, associational approaches \citep[e.g.][]{epsilon-diagnosis} do not aim to provide causal explanations for anomalies, but instead to simply prioritise a small number of variables as potential causes of the anomaly so that they can each be manually investigated by an oncall engineer. The principle is simple: the root cause of an anomaly in a target should have experienced anomalous behaviour at a similar time. In some isolated settings this may be sufficient, the oncall engineer introduces causal information through their domain knowledge and so long as a sufficiently small number of metrics are prioritised for inspection the task is manageable in a timely manner. However, in general this method is unlikely to be suitable. Associational approaches cannot distinguish metrics which are anomalous as a result of the anomaly at the target versus being the root cause of it, nor can they distinguish metrics whose association is due to having a common cause. As such, associational approaches are susceptible to false positives which simply applying a stricter cut-off will not solve.

It is clear that when the goal is to extract \textit{actionable} insight from the data (i.e. to learn which variable needs to be fixed to address the occurrence of the anomaly) a causal approach is required (i.e. at $\mathcal{L}_{2}$ or $\mathcal{L}_{3}$). Although there is not consensus on the way RCA should be characterised as a causal problem, a number of works have proposed formalizations and provided algorithms motivated by them \citep[e.g.][]{budhathoki21a,Budhathoki2022,rca_ikram_neurips,rca_circa}. We assume that the data generating process can be modelled by a Structural Causal Model (SCM) whose underlying causal graph is a directed acyclic graph (DAG). In particular, the SCM describes how each variable $X_{i}$ is generated from its parents $PA_{i}$ in the causal graph and an unobserved noise term $N_{i}$,
$
X_{i} := f_{i}(PA_{i},N_{i}),
$
where $N_1,...,N_{n}$ are jointly independent~\citep{pearl2009}.

Common to all causal approaches to RCA is to treat the occurrence of an anomaly as being the result of a change in the causal mechanism generating the root cause variable. Each method then differs in how to define a root cause and what causal information we have available to us at time of analysis.
At $\mathcal{L}_{2}$ we assume we know the causal graph (or can learn it from data) and treat the mechanism change as a distribution change. In particular, following the causal Markov assumption~\citep{pearl2009}, the joint distribution $P_{\textbf{X}}$ over $(X_{1},...,X_{n})$ factorizes into causal mechanisms
\[
P_{\textbf{X}} = \prod_{i}^{n}P_{X_{i}\mid PA_{i}}.
\]
The change in the joint distribution following the anomaly is then the result of mechanism changes in a subset of variables $X_{T}$, indexed by a change set $T$, where
\[
P_{\textbf{X}}^{T} = \prod_{j\in T}\tilde{P}_{X_{j}\mid PA_{j}}\prod_{j\notin T}P_{X_{j}\mid PA_{j}}
\]
is the joint distribution resulting from the change of causal mechanisms at each $X_{j}$ in $T$ from $P_{X_{j}\mid PA_{j}}$ to $\tilde{P}_{X_{j}\mid PA_{j}}$. A variable $X_{i}$ is then considered a root cause if the change in the joint distribution can be attributed to a change in the causal conditional for $X_{i}$. As a causal contribution problem, the task is to quantify the \textit{extent} to which the change in the joint distribution can be attributed to a change in the causal conditional for a potential root cause~\citep{budhathoki21a}, with root causes being those whose contributions are, in some sense, large. Others reduce the task further and simply call a variable $X_{i}$ a root cause \textit{if} its causal conditional has changed~\citep{rca_ikram_neurips}.

At $\mathcal{L}_{3}$, the finest-grain approach, we assume we know the full SCM (or make suitable assumptions, e.g. additive noise~\citep[e.g.][]{shimizu2006a,hoyer2008a}, such that we can learn it) and the question of whether a variable $X_{i}$ is a root cause is posed counterfactually. For example, \citet{rca_sage} ask "would the anomaly at $X_{n}$ have occurred had $X_{i}$ been at a previously known to be `normal' value?" \citet{Budhathoki2022} refine this further as a causal contribution problem, quantifying the extent to which the counterfactual (tail) probability of the anomalous event $x_{n}$ increases owing to the factual mechanism of $X_{i}$ compared to if it had been as `normal'.

RCA at both layers $\mathcal{L}_{2}$ and $\mathcal{L}_{3}$ is substantially different from determining the behaviour of a target variable $X_{n}$ with respect to an intervention at $X_{i}$. By treating anomalies as the result of a mechanism change, causal RCA asks which variable(s) take on an anomalous value which cannot be explained by their parents. This rules out identifying a variable which simply `transmits' the value of its parent as a root cause.
\section{Related Work} \label{sec:related_work}
In this section, we examine techniques for RCA, split into those specifically targeted at microservice-based applications and those focusing on other use cases, and discuss the content and availability of the datasets on which the methods were evaluated. 

\paragraph{RCA in microservice-based applications}

{\em Pinpoint}~\citep{pinpoint, kiciman2005} is an early example of a method for anomaly detection and RCA in microservice-based applications and is based on detecting changes in the interactions between application components via requests. Pinpoint is also the first method to be evaluated by injecting faults into the Petstore application, upon which our own dataset is based (see section~\ref{sec:description}), however no dataset was made publicly available. Another early method for RCA in microservice architectures is {\em CauseInfer}~\citep{rca_causeinfer} which is based on a traversal of a causal graph between components in a distributed system.
CauseInfer was evaluated on a controlled distributed system of 5 machines injecting issues including a CPU hog, memory leak, disk hog, overload, configuration change, and bugs.  
It has been extended in~\citet{rca_microscope} which was evaluated on a Sock-shop application~\citep{sockshop} where CPU hogs, traffic, and pauses were injected. 
A further extension was presented by \citet{rca_MicroHECL}. This system ranks the anomalous paths by the correlation of the anomaly index between the potential root cause and a target service. It was evaluated on real issues in the e-commerce system of Alibaba.
A similar approach was used in \citet{rca_automap} who present {\em AutoMAP} which performs a random walk on the causal graph
with correlations between potential root causes and the target service serving as edge weights.
It  was evaluated on simulations and on a real-world enterprise cloud system.
Another approach examined using actual operational data from an e-commerce platform (eBay) is the {\em GRANO} system~\citep{grano_ebay}.
None of these datasets are publicly available. 

 A counterfactual approach RCA was presented by \citet{rca_sage} who proposed {\em Sage}.
 Counterfactuals are evaluated by making use of conditional variational autoencoders~\citep{CVAE}. 
Sage is evaluated on a few different microservice architectures developed in DeathStarBench~\citep{DeathStarBench} reflecting a social network, a hotel reservation system, and a media system. However, again these datasets are not publicly available.

More recently, \citet{rca_circa} proposed {\em CIRCA} which models an issue as an intervention on the root cause node.
The method was evaluated in experiments on synthetic data and a banking application.
\citet{rca_ikram_neurips} extended this line of work and proposed a method for identifying root causes in microservice architectures without relying on a service map. 
The method was tested by injecting issues into a Sock-shop application~\citep{sockshop} with data released ~\citet{sockshop_issues}. 
Our dataset complements this dataset and presents new challenges to identify root causes. In our experiments (Sec.~\ref{sec:experiments}) we see that methods that perform well on the Sock-shop data do not necessarily perform well on our dataset. 

\paragraph{RCA for other applications}

\citet{rcaserver}  identify  causes of  performance issues in servers such as a configuration setting. They experiment on Apache, lighttpd, Postfix, and PostgreSQL.
\citet{Budhathoki2022} extend the counterfactual question of~\citet{rca_sage} and account for interactions using Shapely values~\citep{Shapley}. 

They use data on river flows for evaluation.
Broadly related is also explainable AI \citep[see e.g.][]{ribeiro-2016,sundararajan-2017,lundberg-2017} that tries to explain why a machine learning model returned a certain output and which input features were most relevant for this outcome. While most methods aim to explain \emph{any} output of the model, \citet{ide2021anomaly} focus on explaining anomalous outcomes.
They evaluate their approach on a building-energy prediction task. 
These datasets are very different from data collected in applications based on microservice-architectures.

\section{Dataset Description}\label{sec:description}
 In this benchmark, we build on an  application that has been publicly released in~\citet{PetAdoption} and features a pet site for adoptions of different kinds of pets. 
This application composed of microservices  is running on Amazon Web Services (AWS) \citep{AWS}. These microservices include storage (distributed file systems, databases), publish-subscribe systems, load balancers, and custom application logic built in a container and deployed using Kubernetes~\citep{Kubernetes}. 
 The application offers search over pets based on different attributes and permits transactions to adopt and pay for a pet. See the Appendix~\ref{app:application} for a screenshot of the landing webpage.
 Furthermore, the release of the application comes with a traffic generator that we use to simulate user interactions with the website.
 Since our application is build on AWS infrastructure and uses AWS services, in the following, we will refer to them to explain our construction. Note, however, that no prior knowledge or active usage of AWS infrastructure is required to use this dataset. 
 To learn more about the application and its microservices, see~\citet{PetAdoption}.
 We injected artificial performance issues into various microservices.  

\subsection{Data Structure}
The dataset contains multiple scenarios, each with a set of issues to diagnose. 
The \texttt{graph.csv} describes the {\em service map} across microservice nodes obtained from tracing. The service map refers to a graphical representation of the flow of function calls within the system, and it visualizes the interactions between different microservices. 
The metrics at these nodes are recorded in \texttt{metrics.csv} for both a normal period (during which no issues were injected) as well as a range of issues in folders named \texttt{issues0}, \texttt{issues1}, etc. The issues are split evenly at random into \texttt{train} and \texttt{test} sets according to the microservice they originated from. 
This separation allows practitioners to conduct any optimizations for their RCA on \texttt{train} and report performance separately on \texttt{test}.  Each split-folder contains directories with issues. Each issue contains a metrics file and a target file with the ground truth root cause described next. The full directory structure is listed in the Appendix~\ref{app:dataset}.

\subsection{Service Map}
We obtain the service map of the application from AWS X-Ray~\citep{xray} by tracing user requests through the application. An edge from A to B indicates that A calls B. 
Notably, this does not describe the causal graph that connects the metrics originating from those microservices. Nevertheless, it can be useful in constructing a causal graph as noted in~\citet{rca_circa, rca_sage,eulig2023toward}. 
The service map can be loaded from \texttt{graph.csv}, which contains the adjacency matrix indicating the dependencies of the microservices of the application, as follows:
\lstinline!nx.from_pandas_adjacency(pd.read_csv('graph.csv', index_col=0), create_using=nx.DiGraph)!.

\begin{figure}[t]
    \centering
    \includegraphics[width=0.99\textwidth]{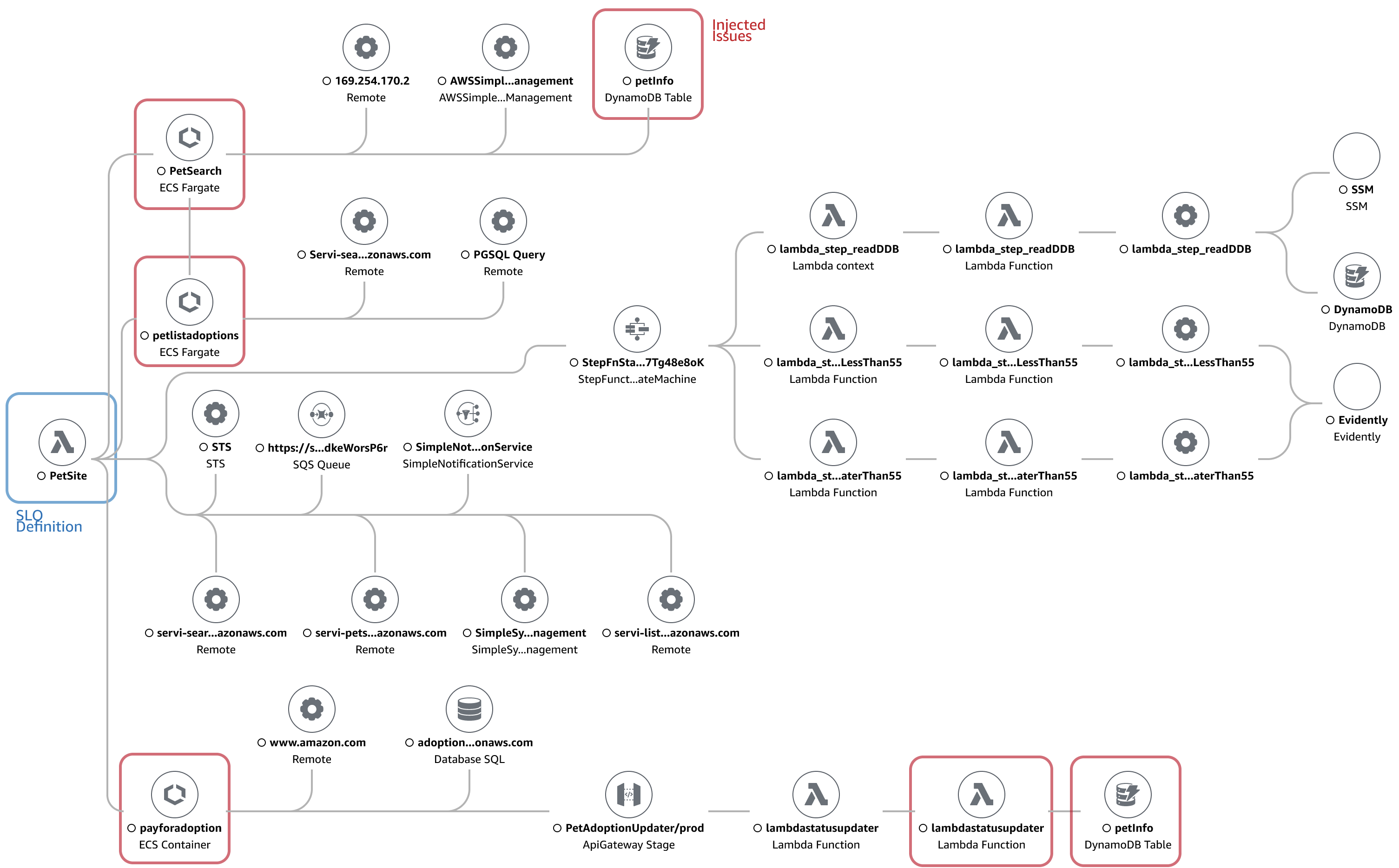}
    \caption{Overview of pet adoption site built on microservices. We injected issues at the highlighted nodes that cause SLO violations of high latencies and low availability at the PetSide node.}
    \label{fig:map}
\end{figure}

Figure~\ref{fig:map} shows the service map of the pet adoption application with directed edges going from left to right. We also highlight nodes in which we injected issues. 

\subsection{Service-level Objectives (SLOs)}
We consider the customer-facing \texttt{PetSite} (see Figure~\ref{fig:map}) as the target node for which we defined service level objectives (SLOs) on what latencies and availabilities are desired to ensure a good end-user experience. For example, a response time SLO at \texttt{PetSite} might be: \enquote{Ensure 95\% of all API requests respond within 200 milliseconds.} This SLO sets a performance target for the response time of an API. It specifies that at least 95\% of all requests should have a response time of 200 milliseconds or less. This means that the system is expected to meet this performance target for the majority of requests, providing a reliable and responsive user experience.
We modified the original code of the application to inject performance issues that lead to violations of such SLOs.

The file \texttt{target.json} contains information about the SLO violation at a target node including the name of the node and the metric as well as a time stamp when the issue was injected. Notably, the SLO violation may happen with a certain delay from this injection times. This delay is due to the time for the change to take effect. Some issues are injected through changes in the Simple System Management service through feature flags, that may be cached locally and polled frequently. 
Furthermore, the file also lists the actual node from which the issue originated and optionally detailing the metric responsible at that node. 
It also contains some additional metadata containing details for reproducibility that can be ignored when determining the root cause.

\begin{example2}
The following describes an example of a SLO violation at the \texttt{PetSite} microservice with an increase in average latency. 
The time of the issue injection is recorded as unix timestamp \texttt{1681390208}, corresponding to 04/13/2023 at 12:50:08pm. 
The correct root cause for this issue is the \texttt{PetSearch} node. No additional information about what happened to cause the issue at that node is given. For this case, the file content is the following:
\end{example2}
\begin{lstlisting}
{
    "target": {
        "node": "PetSite",
        "metric": "latency",
        "agg": "Average",
        "timestamp": 1681390208
    },
    "root_cause": {
        "node": "PetSearch",
        "metric": null
    },
    "metadata": {
        "reproducibility": {
            "command": "aws ssm put-parameter --name '/petstore/searchdelay' --value '500' --overwrite"
        }
    }
}
\end{lstlisting}

\subsection{Traffic Scenarios}
 We generated different traffic patterns using a traffic generator: We have a high-traffic, a low-traffic, and a temporal traffic scenario, in which the traffic varies over the course of a day. 
 
 Under the low-traffic scenario we have on average 485 requests per 5-min, for the high-traffic scenario we have 690 and for the temporal traffic scenario the number of requests vary and average at 571, see Table~\ref{tab:req_per_issue} for quantiles (see also Tables~\ref{tab:latencystats} and \ref{tab:availabilitystats} in Appendix~\ref{app:dataset} for summary statistics of the latency and availability across traffic scenarios, before and after fault injection). 
 
\begin{table}[t]
    \centering
    \caption{We list the scenarios that vary in their request traffic. Here, we report quantiles of the number of requests per 5-minutes.}
    \begin{tabular}{lrrrr}
    \toprule
        Traffic Scenario  & mean & quantile 0.1 & quantile 0.5 & quantile 0.9  \\
        \midrule
         low-traffic & 484 &  464 & 483 & 503 \\
         high-traffic & 690 & 668 &  688 & 714 \\
         temporal-traffic & 571 & 376 & 497 &  884 \\
         \bottomrule
    \end{tabular}
    \label{tab:req_per_issue}
\end{table}

\subsection{Metrics}
\begin{figure}[t]\vspace{0.3cm}
    \centering
    \includegraphics[width=0.99\textwidth]{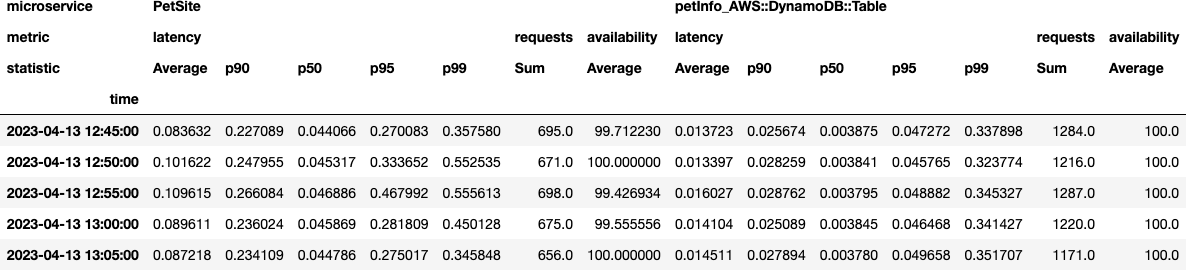}
    \caption{Exemplary printout of the metrics for two microservices with 5 measurements each. The multi-column format contains the node, the metric, and different statistics for the aggregation of the metric over 5 minute time windows.}
    \label{fig:metrics}
\end{figure}

In our pet adoption application we collect the number of requests, their latency (average and different quantiles) and average availability at each microservice always over 5 minute windows. Here, availability is defined as $1 - \frac{\# errors}{\# requests}$. We obtained these metrics from Amazon CloudWatch~\citep{cloudwatch}, a tool to collect and analyze resource and application data. 
Figure~\ref{fig:metrics} illustrates the metrics for two exemplary nodes and five time stamps. 
The \texttt{metrics.csv} file can be loaded using pandas with \lstinline!pd.read_csv('metrics.csv', header=[0, 1, 2], index_col=0)!.


\subsection{Injected Issues}
We list out the different issues we injected below, including two of the most common issues in cloud systems: memory leaks and CPU hogs~\citep[see][]{faultsinservices, rca_ikram_neurips}, and provide a summary in Appendix~\ref{app:dataset} (Table ~\ref{tab:issuebreakdown}). We parameterized many of these issues by the time delay or the random fraction of traffic for which they would occur. Further, we added AWS System Manager parameters to many of the nodes  that we can modify on the fly to turn on and off issues of varying severity without requiring a deploy. 
Overall, we made sure the error percent or delay in milliseconds were large enough to lead to noticeable spikes on the \texttt{PetSite} for at least some traffic scenarios. 
We repeated each issue twice so that we can study variability. Overall, we also cover a wide range of anomaly severities, from mild deviations from normality ($\sim$1-2 standard deviation events) to rare extreme events (>3 standard deviations) (see Table~\ref{tab:itscorestats} and Figures~\ref{fig:itlatencyhist} and \ref{fig:itavailabilityhist} in Appendix~\ref{app:dataset} for summary statistics and the distributions IT anomaly scores for latency and availability metrics at \texttt{PetSite}). This will enable practitioners at companies to see how closely the severities of issues explored in our dataset match those they observe during issues, and determine whether or not our work is relevant for their setting.

 \textbf{Requests Overload:}
 We overload the database with external requests hitting the throttling limit. This causes issues in the \texttt{petInfo DynamoDB Table} node marked in Figure~\ref{fig:map}.
This issue is mimics a real issue we encountered.
 
 \textbf{Memory Leak:}
 The application comes with a memory leak that may trigger for requests of a specific type (concerning the adoption of bunnies only). This causes errors in the \texttt{payforadoption ECS Fargate} service that propagates to the \texttt{PetSite}.
 Memory leaks are common issues in cloud systems~\citep[see][]{faultsinservices, rca_ikram_neurips}.
 
 \textbf{CPU hog:}
 We introduce a CPU hog in the \texttt{lambdastatusupdater Lambda Function} that keeps the CPU busy for a certain duration of time through multiplications of random numbers, which leads to an increased latency that propagates to the \texttt{PetSite}. CPU hogs  represent common issues in cloud systems~\citep[see][]{faultsinservices, rca_ikram_neurips}.

 \textbf{Misconfiguration:}
 The application comes with a misconfigured storage bucket in the \texttt{PetSearch ECS Fargate} microservice. This reflects a common problem of insufficient permissions. We randomly select 1\% - 2\% of requests subject to this error.
We additionally trigger a misconfigured database table name in the \texttt{lambdastatusupdater Lambda Function} and the \texttt{petlistadoptions ECS Fargate} microservice. 
 These misconfigurations lead to a decrease in availability of \texttt{PetSite}. For the \texttt{lambdastatusupdater Lambda Function} this error affects requests for kittens. For the \texttt{petlistadoptions ECS Fargate} microservice the percentage of requests that used an invalid database table ranges from 2\% to 10\%. 
 
\begin{figure}[t]
    \centering
    \includegraphics[width=0.65\textwidth]{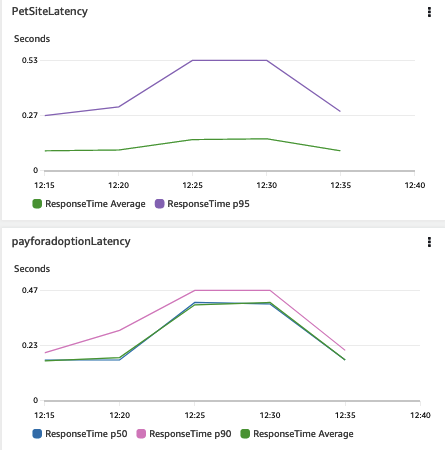}
    \caption{Example showing how delays in the \texttt{payforadoption ECS Fargate} microservice reflect in the response times at the PetSite.}
    \label{fig:delay}
\end{figure}

 \textbf{Other Delays:}
 We also introduced delays in \texttt{PetSearch ECS Fargate} and \texttt{payforadoption ECS Fargate} through sleep statements. An example for such a delay and its propagation to \texttt{PetSite} is shown in Figure~\ref{fig:delay}. 
These naive sleep statements are meant to model performance regressions of microservices that can be introduced by code changes.
 
To instrument delays in \texttt{PetSearch ECS Fargate} we introduced a Systems Manager parameter that governs the delay in milliseconds (between 500 and 2000) that is injected for search requests of bunnies. 
 In order to limit the impact from retrieving this parameter we implemented a caching layer of the parameters refreshing them once a minute. That way we limit requests to the Systems manager while guaranteeing some freshness. As a consequence though there can be some time that passes between the trigger time (when we turn on the parameter) and the service refreshes the parameter. 
The fact that the delay only impacts the bunny requests means that its not all latency percentiles are effected the same way. 
To instrument delays in \texttt{payforadoption ECS Fargate}  we introduced a Systems Manager parameter that governs the delay in milliseconds (between 250 and 1000) that is injected for all requests. 
  
In total we have 68 issues triggered at 5 different nodes across the three traffic scenarios causing SLO violations due to high latency or low availability at the PetSite node. For each of the 68 issues, there is exactly one ground-truth root cause, based on where we inject the issue. Table~\ref{tab:num_issues} lists how they are split across the three traffic scenarios and latency and availability performance measures.

\begin{table}[t]
    \centering
    \caption{We list the number of issues affecting the latency and availability at the PetSite node for each of the distinct traffic scenarios.}
    \begin{tabular}{lrr}
    \toprule
        Traffic Scenario  & number of latency issues & number of availability issues \\
        \midrule
         low-traffic & 14 & 12 \\
         high-traffic & 14 & 12 \\
         temporal-traffic & 8 & 8 \\
         \bottomrule
    \end{tabular}
    \label{tab:num_issues}
\end{table}

\subsection{Evaluation}
A method for RCA can take as input the service map, the metrics, and information about the SLO violation (including the target node and metric) and needs to output a list of potential root causes composed of a node and potentially a metric with a confidence score. This could look as follows:

\begin{lstlisting}[language=Python]
def analyze_root_causes(graph: nx.DiGraph, target_node: str, target_metric: str, target_statistic: str, normal_metrics: pd.DataFrame, abnormal_metrics: pd.DataFrame) -> List[PotentialRootCause]:
    """Finds root causes of a performance issue in target_node."""
\end{lstlisting}

Using these outputs, we provide a method to evaluate the given RCA approach on the dataset. We compute the top-1 and top-3 recall. Given that there is a unique root cause that represents the ground truth the top-1 recall captures the accuracy of the method to determine the correct root cause. In practice, a method can be used to present an oncall engineer with a ranked list of root-causes. The engineer can then investigate further. To capture the quality of this ranked list, we additionally compute the top-3 recall. An approach can be evaluated calling the following method: 

\begin{lstlisting}[language=Python]
def evaluate(analyze_root_causes, dir: str, split: str=None) -> pd.DataFrame:
    """Computes the top-k recall of the method to analyze root causes."""
\end{lstlisting}
\section{Experiments}\label{sec:experiments}

We provide an initial comparison of proposed RCA methods on the 68 injected issues.

\noindent\textbf{Methods}

We test both methods relying on the causal graph and those that do not. 
The causal formalization of RCA makes clear that we would expect methods which are given causal information, in the form of the causal graph, to perform better than those which are either not causal (owing to false positives) or which must learn the causal graph or SCM from the data (owing to needing to make strong assumptions and the statistical challenge of learning with limited data). However, as in practice we often do not know the causal graph, we also assess methods that do not require it. 

\noindent\textbf{Methods that require a causal graph}
Inspired by \citet{rca_causeinfer, rca_microscope, rca_MicroHECL}, the \emph{traversal} method identifies a node as a root cause under two conditions: 1) none of its parent nodes exhibits anomalous behavior, and 2) it is linked to the SLO violating node exclusively through nodes showing anomalous behavior. This thereby encodes the requirement that the anomalous behaviour at a root cause should not be explained by its parents, as follows from the formalism that an anomaly occurs due to a mechanism change at the root cause, and that it should contribute to the anomalous behaviour at the target. Although simple, traversal is therefore an $\mathcal{L}_{2}$ method. 
For labeling a node $X$ as anomalous based on a given observation $x$, we use the MAD-score with a threshold of 5. This score is the normalized distance to the median defined as $\frac{|x - \text{median}[X]|}{\text{MAD}[X]}$, where $\text{MAD}[X]$ denotes the median absolute deviation. We have experimented with other anomaly scores implemented within the Python library {\em DoWhy}~\citep{dowhygcm} that we describe in the Appendix~\ref{app:anomaly} from which MAD-score performed best.

\emph{CIRCA}~\citep{rca_circa} likewise requires a causal graph, and identifies root causes by testing which causal conditional distributions have changed. Practically this is achieved by fitting a linear Gaussian SCM to the data from the normal period, and comparing the predicted value of each variable given its parents to the true value in the abnormal period. CIRCA is therefore does not treat RCA counterfactually and is an $\mathcal{L}_{2}$ method.

\emph{Counterfactual Attribution}~\citep{Budhathoki2022}, like CIRCA, requires an input causal graph which is then used to fit an SCM, but instead treats RCA as a counterfactual contribution problem (as described in Sec. \ref{sec:background}), returning Shapley value-based contribution scores for each potential root cause. This is therefore an $\mathcal{L}_{3}$ method.

\noindent\textbf{Methods that do not require a causal graph}
Without a causal graph we must either learn one from the data or else rely on associational, $\mathcal{L}_{1}$, methods. A simple heuristic used in practice is to rank potential root causes according to their \emph{correlation} with the target in the abnormal period. In this study we filter for nodes which have been detected as anomalous using the MAD-score, use the Pearson correlation coefficient and rank according to its \textit{p}-value as a simple baseline.

\emph{$\epsilon$-Diagnosis}~\citep{epsilon-diagnosis} constructs a test statistic based on the energy distance correlation coefficient~\citep{székely_rizzo_2014} and performs a two-sample test to identify which services changed significantly from the normal to abnormal periods. As this does not assess whether the change in a service can be explained by its parents, this is a $\mathcal{L}_{1}$ method.

\emph{RCD}~\citep{rca_ikram_neurips} exploits the fact that mechanism changes can be modelled as soft interventions, and thereby applies work on causal discovery with unknown interventions~\citep{jaber2020} to identify root causes. More precisely, by introducing a binary `intervention indicator' variable, testing which variables' causal conditional distributions have changed between the normal and abnormal periods amounts to checking conditional independence statements involving the indicator: for the variables which cannot be made conditionally independent of the intervention indicator it must also be the case that their `anomalousness' cannot be explained by their parents. These variables are therefore the children of the intervention indicator in an extended causal graph, which can be identified under faithfulness and causal sufficiency. RCD is therefore an $\mathcal{L}_{2}$ method.

For evaluation we use the implementation by~\citet{PyRCA} for the methods CIRCA, $\epsilon$-Diagnosis and RCD and the implementation by~\citep{dowhygcm} for the counterfactual attribution method. We opensource (link to be added upon acceptance) implementations of the traversal and the correlation methods.
In the Appendix~\ref{app:methods} we describe the configuration options we tested for each method and which one performed best and is used in the results below.

\noindent\textbf{Results.} As expected the results presented in Table~\ref{tab:resultsnew} illustrate that methods with access to a causal graph typically perform better than those without. However, counterfactual attribution struggles at identifying root causes for drops in availability. This is likely due to the low variability in availability across the normal period, making learning an SCM challenging. 
By comparison, as traversal needs only to perform anomaly detection, it is robust to low variability in the normal period and to small numbers of data points in the abnormal period. 

When it comes to methods that do not have access to the graph, however, the performance drops dramatically. In fact, none of the proposed approaches outperform the basic ranked correlations. 
This calls for more research into the development of robust causal attribution methods which to not depend on knowing the causal graph. 
While RCD and $\epsilon$-diagnosis achieved good performances on hundreds of data-points (e.g. 1 second aggregated metrics) as illustrated by~\citet{rca_ikram_neurips} they are not designed to handle limited data as present in this dataset. It is noteworthy, that in the top-1 recall (see Table~\ref{tab:resultsnewtop1} in Appendix~\ref{app:results}) ranked correlation outperforms all other methods including those with access to the graph. We hypothesize that both the data size and the graph structure play a role here. Pairwise-correlations work better with little data compared to estimating SCMs or discovering graph structures. Additionally, in these experiments we focus on SLO violations at the PetSite node, which, in the causal graph, is a leaf and does not share a common cause with another node. As a result false positives arising from being unable to distinguish its causes and effects, and association due to common causes, are not an issue. We hope future work sheds more light on error sources (including errors arising from finite samples, and causal graph misspecification). To investigate whether method performance would vary according to the severity of the injected issue, we additionally evaluated performance stratified by IT anomaly score (see Appendix~\ref{app:anomaly}), and find that most methods do not consistently perform better across both latency and availability issues when the strength of the observed anomaly is larger (see Tables~\ref{tab:availabilitytop3byscore} and \ref{tab:latencytop3byscore}).

\begin{table}[t]
    \centering
    \caption{Top-3 recall of the RCA methods measuring the accuracy of including the correct root-cause node in the top-3 results.}
    \begin{tabular}{ll|ccc|ccc}
    \toprule
    & & \multicolumn{3}{c|}{graph given}  & \multicolumn{3}{c}{graph not given} \\
    \midrule
traffic & metric & 
traversal & circa & counter-  &  $\epsilon$-diagnosis  & rcd & correlation
\\
scenario & & & & factual & & \\
\midrule
low  & latency & 
0.57 & \textbf{0.86} & 0.71 & 0.00 & 0.21 & 0.57\\
low  & availability & 
\textbf{1.00} & \textbf{1.00} & 0.42 & 0.00 & 0.75 & 0.92\\
high  & latency & 
0.79 & \textbf{1.00} & 0.86 & 0.00 & 0.07 & 0.79\\
high  & availability & 
\textbf{1.00} & 0.00 & 0.00 & 0.33 & 0.00 & 0.92\\
temporal  & latency & 
\textbf{1.00} & \textbf{1.00} & 0.50 & 0.12 & 0.75 & 0.75\\
temporal  & availability & 
\textbf{1.00} & \textbf{1.00} & 0.25 & 0.12 & 0.75 & 0.75\\
         \bottomrule
    \end{tabular}
    \label{tab:resultsnew}
\end{table}

Additionally, we tested the specificity on normal data but found that all methods fabricate root-causes during normal operations. We leave the adaption of the methods for improved specificity for future work. Setting a threshold on the scores can be a first starting point. 

\section{Limitations}\label{sec:limitations}

This dataset is generated from an Web application running on cloud infrastructure created for demonstration. 
It is not an application running in production with real user traffic. The traffic is generated artificially.
This is a crucial limitation in our dataset. 
In real applications it is possible that issues affect user behavior. In case of errors or increased latencies users may refresh a page, restart an application or abandon their session. These reactions lead to a change in request which in turn can affect metrics across all microservices of an application. 
As such, causal approaches which assume the underlying causal graph is acyclic 
may perform well on our benchmark but not in the presence of such feedback loops. Work on extending causal formalizations of RCA to non-recursive models is an important future direction.

Additionally, we do not consider cases where there is more than one root cause as discussed in \citet{oesterle2023beyond}. While two services independently failing at once is rather unlikely, multiple root causes can still occur in practice e.g.\ due to failures at unmeasured confounders. 

Further, the issues we have injected into microservices are likely to be different from issues encountered in real-world applications. While they include some common issues in cloud systems (i.e. memory leaks and CPU hogs)~\citep[see][]{faultsinservices, rca_ikram_neurips} in addition to request overloads and misconfigurations, they lack coverage and are not representative. Furthermore, we performed some tuning of the parameters for issue injection (the fraction of requests to be affected or the delay in ms) to make sure that they have a pronounced effect in the system, and thereby introduce a bias for stronger anomalous events. We might expect therefore that the dataset not give a complete picture of method performance for weak anomalous events, however, when the evaluation is stratified by anomaly severity, we do not see that methods perform consistently better for stronger anomalies (see Tables~\ref{tab:availabilitytop3byscore} and \ref{tab:latencytop3byscore} in Appendix~\ref{app:results}).

Lastly, the methods we experimented with have not been tuned or massaged. That way they give us a sense for the performance they can give out-of-the-box for this new dataset. Moreover, we have not included all RCA methods in the comparison.

\section{Conclusion}\label{sec:conclusion}
In this paper, we introduced a dataset that encompasses metric data from a microservices-based application during both normal and anomalous operational periods. We believe that this dataset could serve as a valuable resource for evaluating and benchmarking RCA techniques. For illustration, we report the top-3 recall for a selection of published RCA methods. These  methods approach the problem of defining a root cause following different formalizations spanning the causal hierarchy~\citep{Bareinboim2022-BAROPH-2}.
We find that causal methods perform well when we provide the causal graph, but that methods relying on learning the causal graph or the full SCM do not yield satisfactory performance when data is limited. In this case the simple baseline of ranking potential root causes according to the \textit{p}-value for correlation with the target, and filtering for variables detected as anomalous, provides a strong baseline against which new causal methods should be evaluated. We thus hope that with this dataset we can enable future research into robust methods for root cause analysis that work well with limited data. In order to work in practice, such methods should not require access to the causal graph and should be able to work with a handful of abnormal measurements to make a timely diagnosis.

We anticipate that this dataset will inspire further exploration and advancement in the field of RCA methods. Additionally, we encourage the community to contribute by expanding this dataset to cover other applications and supplement it with performance issues observed in real-world production systems.

\newpage
\bibliography{references}

\newpage
\appendix
\section{Application}\label{app:application}
\begin{figure}[h]
    \centering
    \includegraphics[width=0.99\textwidth]{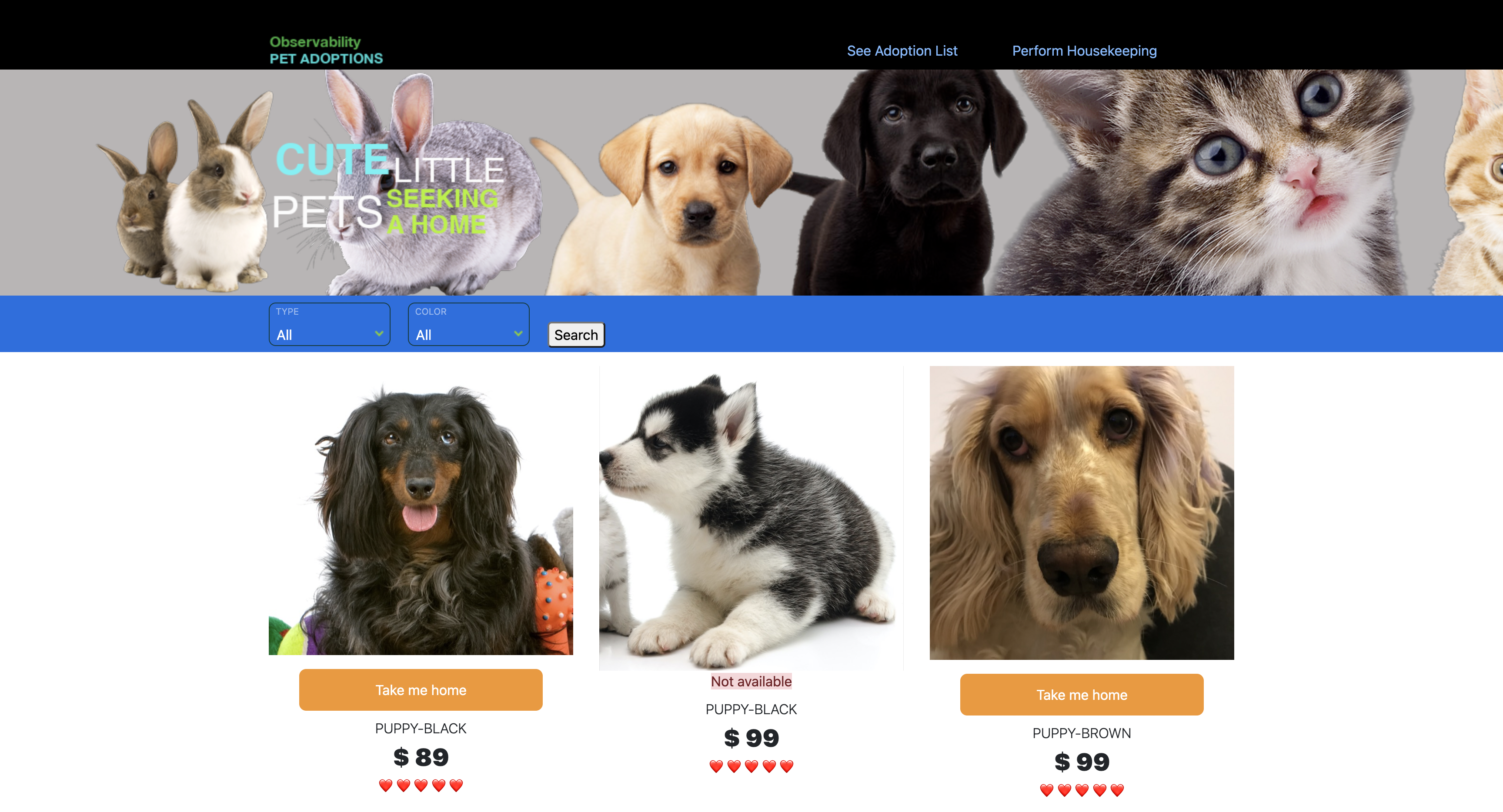}
    \caption{Screenshot of the landing webpage of the pet adoption application.}
    \label{fig:petshop}
\end{figure}

 Figure~\ref{fig:petshop} shows a screenshot of the pet adoption  webpage. It supports search and purchasing features powered by different microservices.

\section{Dataset}\label{app:dataset}

The data is stored according to the following structure:
\begin{lstlisting}
scenario/
    graph.csv
    noissue/
        metrics.csv
    train/
        issue0/
            metrics.csv
            target.json
        issue1/
            metrics.csv
            target.json
        ...
    test/
        issue0/
            metrics.csv
            target.json
\end{lstlisting}

Table~\ref{tab:issuebreakdown} gives a detailed breakdown of each of the injected issues.
{
\renewcommand{\arraystretch}{3}
\begin{sidewaystable}
\centering
\begin{adjustbox}{width=\textwidth,center=\textwidth}
\begin{tabular}{c|m{3cm}|c|c|m{4cm}|c|c|c|m{4cm}}
\toprule
\textbf{Node} & \textbf{Type of Node} & \textbf{Target Metric} & \textbf{Type of Issue} & \textbf{Relevance} & \textbf{Traffic Impacted} & \textbf{Parameter Range} & \textbf{Total Count} & \textbf{Description} \\ \hline
petInfo & database & availability & traffic spike / throttling & as practitioners we have run into this type of issue & a fraction of traffic will be throttled & - & 6 & We generate many read requests directly hitting the database, triggering throttling. \\ \hline
petSearch & containerized micro-service on ECS & availability & human configuration error & configurations are often done manually and cannot be tested as easily as code & random fraction of all traffic & 1\%, 2\%, 3\% & 10 & A random fraction of requests attempts to access a directory in the file system that does not exist. \\ \hline
petSearch & containerized micro-service on ECS & latency & performance degradation & performance degradations can happen due to deploying less performant code, change in hardware, overall resource exhaustion & bunny requests & 0.5sec, 1sec, 2sec & 10 & Bunny requests are delayed by some number of seconds. \\ \hline
payforadoption & containerized micro-service on ECS & availability & memory leak & among the most common issues in cloud systems, see \cite{faultsinservices, rca_ikram_neurips} & bunny requests & n/a & 6 & A memory leak is triggered for all requests. \\ \hline
payforadoption & containerized micro-service on ECS & latency & performance degradation & performance degradations can happen due to deploying less performant code, change in hardware, overall resource exhaustion & all & 0.25sec, 0.5sec, 1sec & 10 & All requests are delayed by some number of seconds. \\ \hline
statusupdater & serverless data-processing system on lambda & availability & human configuration error & configurations are often done manually and cannot be tested as easily as code & kitten requests & n/a & 6 & Kitten requests attempt to access a database table that does not exist. \\ \hline
statusupdater & serverless data-processing system on lambda & latency & performance degradation & performance degradations can happen due to deploying less performant code, change in hardware, overall resource exhaustion & all requests & 0.25sec, 0.5sec, 1sec & 10 & All requests are delayed by some number of seconds. \\
\bottomrule
\end{tabular}
\end{adjustbox}
\caption{Breakdown of injected issues. In all cases, issues were repeated twice.}
\label{tab:issuebreakdown}
\end{sidewaystable}
}

\begin{table}[h]
\centering
\begin{tabular}{ccccc}
\toprule
traffic scenario & condition & mean (ms) & std &  \\ \hline
low                       & normal             & 9.49               & 0.57         &  \\
low                       & abnormal           & 13.46              & 8.53         &  \\
high                      & normal             & 9.82               & 0.42         &  \\
high                      & abnormal           & 14.42              & 11.38        &  \\
temporal                  & normal             & 10.04              & 0.90         &  \\
temporal                  & abnormal           & 24.10              & 21.34        &  \\
\bottomrule
\end{tabular}
\caption{Mean and standard deviations of latency across traffic scenarios and normal and abnormal conditions.}
\label{tab:latencystats}
\end{table}

\begin{table}[h]
\centering
\begin{tabular}{ccccc}
\toprule
traffic scenario & condition & mean (\%) & std &  \\ \hline
low                       & normal             & 99.74              & 2.48         \\
low                       & abnormal           & 98.60              & 1.85         \\
high                      & normal             & 99.91              & 0.23         \\
high                      & abnormal           & 98.66              & 1.80         \\
temporal                  & normal             & 99.85              & 0.95         \\
temporal                  & abnormal           & 98.19              & 2.37         \\
\bottomrule
\end{tabular}
\caption{Mean and standard deviations of availability across traffic scenarios and normal and abnormal conditions.}
\label{tab:availabilitystats}
\end{table}

\begin{table}[h!]
\centering
\begin{tabular}{ccccc}
\toprule
traffic scenario & metric & min & max &  \\ \hline
low                       & latency         & 1.05               & 7.07               \\
low                       & availability    & 2.70               & 5.97               \\
high                      & latency         & 3.71               & 7.07               \\
high                      & availability    & 1.52               & 7.07               \\
temporal                  & latency         & 5.27               & 8.10               \\
temporal                  & availability    & 4.25               & 5.53               \\
\bottomrule
\end{tabular}
\caption{Minimum and maximum IT anomaly scores across each traffic scenario and metric.}
\label{tab:itscorestats}
\end{table}

\begin{figure}[h!]
    \centering
    \includegraphics[width=\textwidth]{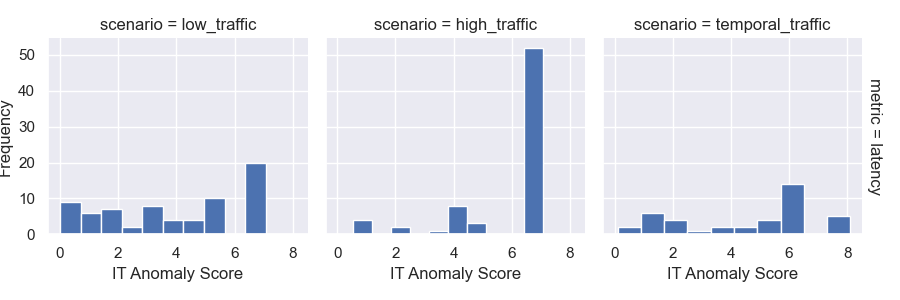}
    \caption{Distribution of IT anomaly scores at PetSite for latency issues.}
    \label{fig:itlatencyhist}
\end{figure}

\begin{figure}[h!]
    \centering
    \includegraphics[width=\textwidth]{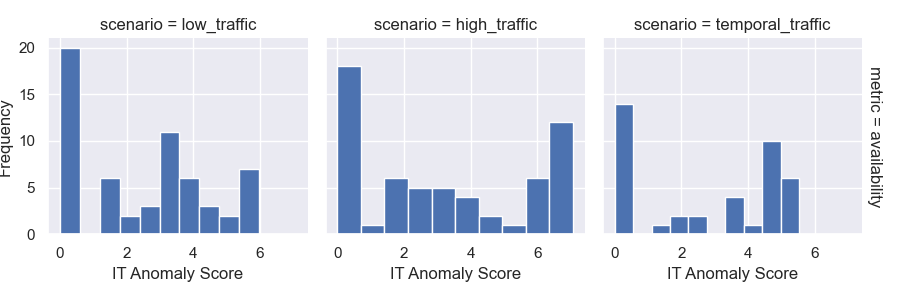}
    \caption{Distribution of IT anomaly scores at PetSite for availability issues.}
    \label{fig:itavailabilityhist}
\end{figure}

\section{Experiments}

\subsection{Anomaly Scores}\label{app:anomaly}
For the anomaly traversal we experimented with the following anomaly scores:
\begin{itemize}
    \item \textbf{z-score:} This is a measure of the normalized distance from the mean, calculated as $\frac{|x - \mathbb{E}[X]|}{\sqrt{\text{Var}[X]}}$.
    \item \textbf{MAD-score:} This score is the normalized distance to the median defined as $\frac{|x - \text{median}[X]|}{\text{MAD}[X]}$, where $\text{MAD}[X]$ denotes the median absolute deviation.
    \item \textbf{IT-score:} This approach translates the scores defined above into an information theoretic quantity, as per \cite{Budhathoki2022}, $-\text{log}(g(X) \geq g(x))$. Here, $g$ is a 'feature' map that functions as an anomaly scorer, producing a score such as the z-score or MAD-score.
    \item \textbf{Median quantile-score:} This score resembles the log-probability of the 2-sided quantile calculated by $-\text{log}(2 \cdot \text{min}\{\text{Pr}[X >= x], \text{Pr}[X <= x]\})$.
\end{itemize}

\subsection{Methods}\label{app:methods}
We experimented with a few options for the various methods and report the best results. 
In particular, for RCD we ran it with the global option, on all metrics, and with a mean imputation scheme. Best worked the interpolation as imputation method that we report in the results.
For $\epsilon$-diagnosis we also tried setting the significance level $\alpha$ to 0.01, using interpolation, and selecting only metrics of the same type as the target metric (e.g. latency or availability) vs using both. Using both performed better and is shown in the results. 
For CIRCA, we also tried both mean and interpolation for the imputation, both max and sum aggregation, and with and without the descendant adjustment. We include the best performing setting with the mean imputation and the max aggregation in the results.

For correlation we experimented with using the coefficient for ranking and/or the \textit{p}-value.

The counterfactual method had the longest run-time owing to the method's reliance on Shapley values.

\subsection{Results}\label{app:results}
Table~\ref{tab:resultsnewtop1} shows top-1 recall for identifying the root cause. Tables~\ref{tab:availabilitytop3byscore} and \ref{tab:latencytop3byscore} show the top-3 recall for each method stratified according the strength of the anomaly detected at the target (PetSite). In particular, small anomalies are those with an IT anomaly score (see Appendix~\ref{app:anomaly} above) lower than the median score across issues of the same type (either latency or availability), whereas large anomalies are those with a score greater than the median.
\begin{table}[h]
    \centering
    \begin{tabular}{ll|ccc|ccc}
    \toprule
    & & \multicolumn{3}{c|}{graph given}  & \multicolumn{3}{c}{graph not given} \\
    \midrule
traffic & metric & 
traversal & circa & counter-  &  $\epsilon$-diagnosis  & rcd & correlation
\\
scenario & & & & factual & & \\
\midrule
low  & latency & 
\textbf{0.57} & 0.36 & 0.36 & 0.00 & 0.07 & 0.43\\
low  & availability & 
0.50 & 0.42 & 0.00 & 0.00 & 0.58 & \textbf{0.75}\\
high  & latency & 
0.57 & 0.50 & 0.57 & 0.00 & 0.00 & \textbf{0.64}\\
high & availability & 
0.33 & 0.00 & 0.00 & 0.00 & 0.00 & \textbf{0.83}\\
temporal  & latency & 
\textbf{1.00} & 0.75 & 0.38 & 0.12 & 0.25 & 0.62\\
temporal  & availability & 
0.38 & 0.38 & 0.00 & 0.00 & 0.50 & \textbf{0.62}\\
         \bottomrule
    \end{tabular}
        \caption{Top-1 recall of the RCA methods measuring the accuracy of identifying the correct root-cause node.}
    \label{tab:resultsnewtop1}
\end{table}

\begin{table}[t]
	\centering
	\begin{tabular}{ll|ccc|ccc}
	\toprule
	& & \multicolumn{3}{c|}{graph given}  & \multicolumn{3}{c}{graph not given} \\
	\midrule
traffic & anomaly & traversal & circa & counter-  &  $\epsilon$-diagnosis  & rcd & correlation \\
scenario & size & & & factual & & \\
\midrule
low & small & 
\textbf{1.00} & \textbf{1.00} & 0.25 & 0.00 & \textbf{1.00} & 0.88\\
low & large & 
\textbf{1.00} & \textbf{1.00} & 0.75 & 0.00 & 0.25 & \textbf{1.00}\\
high & small & 
0.75 & 0.00 & 0.00 & 0.25 & 0.00 & \textbf{1.00}\\
high & large & 
0.75 & 0.00 & 0.00 & 0.38 & 0.00 & \textbf{0.88}\\
temporal & small & 
\textbf{1.00} & \textbf{1.00} & 0.33 & 0.17 & 0.67 & 0.67\\
temporal & large & 
\textbf{1.00} & \textbf{1.00} & 0.00 & 0.00 & \textbf{1.00} & \textbf{1.00}\\
		\bottomrule
	\end{tabular}
		\caption{Top-3 recall for methods for availability issues stratified by the size of the anomaly at the target. Small: IT anomaly score < median score for availability issues, large: IT anomaly score > median for availability issues.}
		\label{tab:availabilitytop3byscore}
\end{table}

\begin{table}[t]
	\centering
	\begin{tabular}{ll|ccc|ccc}
	\toprule
	& & \multicolumn{3}{c|}{graph given}  & \multicolumn{3}{c}{graph not given} \\
	\midrule
traffic & anomaly & traversal & circa & counter-  &  $\epsilon$-diagnosis  & rcd & correlation \\
scenario & size & & & factual & & \\
\midrule
low & small & 
0.40 & \textbf{1.00} & 0.80 & 0.00 & 0.20 & 0.20\\
low & large & 
0.67 & \textbf{0.78} & 0.67 & 0.00 & 0.22 & \textbf{0.78}\\
high & small & 
0.67 & \textbf{1.00} & 0.83 & 0.00 & 0.00 & 0.50\\
high & large & 
0.88 & \textbf{1.00} & 0.88 & 0.00 & 0.12 & \textbf{1.00}\\
temporal & small & 
\textbf{1.00} & \textbf{1.00} & 0.50 & 0.00 & 0.67 & 0.67\\
temporal & large & 
\textbf{1.00} & \textbf{1.00} & 0.50 & 0.50 & \textbf{1.00} & \textbf{1.00}\\
		\bottomrule
	\end{tabular}
		\caption{Top-3 recall for methods for latency issues stratified by the size of the anomaly at the target. Small: IT anomaly score < median score for latency issues, large: IT anomaly score > median for latency issues.}
		\label{tab:latencytop3byscore}
\end{table}
\end{document}